\documentclass[twocolumn,aps,prd,nofootinbib]{revtex4-2}
\usepackage[colorlinks, linkcolor={blue},citecolor={red}]{hyperref}
\usepackage{graphicx}
\usepackage{epstopdf}
\usepackage{amsmath,amssymb,amsfonts}
\usepackage{cleveref}
\usepackage{xcolor}
\usepackage{braket}
\usepackage{esvect}
\newcommand{\ba}{\begin{eqnarray}} 
\newcommand{\ea}{\end{eqnarray}}

\newcommand{\be}{\begin{equation}}
\newcommand{\ee}{\end{equation}}	

\newcommand{\bi}{\begin{itemize}}
\newcommand{\ei}{\end{itemize}}	

\makeatletter

\newcommand*\rel@kern[1]{\kern#1\dimexpr\macc@kerna}
\newcommand*\widebar[1]{%
	\begingroup
	\def\mathaccent##1##2{%
		\rel@kern{0.8}%
		\overline{\rel@kern{-0.8}\macc@nucleus\rel@kern{0.2}}%
		\rel@kern{-0.2}%
	}%
	\macc@depth\@ne
	\let\math@bgroup\@empty \let\math@egroup\macc@set@skewchar
	\mathsurround\z@ \frozen@everymath{\mathgroup\macc@group\relax}%
	\macc@set@skewchar\relax
	\let\mathaccentV\macc@nested@a
	\macc@nested@a\relax111{#1}%
	\endgroup
}

\makeatletter
\renewcommand\@makefnmark{\hbox{\@textsuperscript{\normalfont\color{red}\@thefnmark}}}
\makeatother
\input{epsf.sty}

\begin{document}
\title{Black Hole Remnants from Dynamical Dimensional Reduction?}
\author{Frank Saueressig}
\email{f.saueressig@science.ru.nl}
\author{Amir Khosravi}
\email{AmirPouyan.khosravikarchi@ru.nl}
\affiliation{Institute for Mathematics, Astrophysics and Particle Physics (IMAPP),Radboud University Nijmegen, Heyendaalseweg 135, 6525 AJ Nijmegen, The Netherlands}

\begin{abstract}
 A intriguing feature shared by many quantum gravity programs is the dynamical decrease of the spectral dimension from $D_s = 4$ at macroscopic to $D_s \approx 2$ at microscopic scales. In this note, we study the impact of this transition on the energy loss of static, spherically symmetric black holes due to Hawking radiation. We demonstrate that the decrease in the spectral dimension renders the luminosity of a black hole finite. While this slightly increases the life-time of light black holes, we find that this mechanism is insufficient to generate long-lived black hole remnants. We briefly comment on the relation of our findings to previous work on this topic.
\end{abstract}
\maketitle

\section{Introduction}
Black holes provide a powerful laboratory for testing our ideas about space and time at both the theoretical as well as the observational frontier. A striking theoretical prediction based on quantum field theory in curved spacetime is that black holes are not entirely black: their event horizon emits black body radiation with a temperature inversely proportional to the mass of the black hole \cite{Hawking:1975vcx,haw1,haw2,haw3}. Owed to this so-called Hawking effect, the black hole loses mass and eventually evaporates completely within a finite time-span. The robustness of this scenario has been corroborated in a number of different ways comprising perturbative computations in a fixed background spacetime \cite{Hawking:1975vcx}, the detector approach \cite{detector0,detector00,detector,detector1}, as well as by analogy to the Unruh effect \cite{Unruh:1976db}.

The Hawking effect gives rise to a series of theoretical puzzles though. Firstly, one encounters the black hole information paradox reviewed in \cite{chen}. The picture above suggests that the physical information about how the black hole was formed could permanently disappear, allowing many physical states to evolve into the same state. Basically, there are three viewpoints on this problem \cite{stev1,dan}: 1) information is indeed lost after the black holes has evaporated completely, 2) evaporation stops and information is preserved inside a stable remnant, and 3) information may be returned outside via Hawking radiation. In particular, with regard to option 2) it is important to understand the final configuration emanating from the black hole evaporation process. 

The second puzzle comes from the increase of the horizon temperature as the black hole becomes lighter and lighter. For instance, a black hole with a mass of the order of the Planck mass would have a temperature $T_h \approx 10^{31}$K. Theoretically, it is then predicted that the final stage of the black hole evaporation process generates a so-called thunderbolt singularity. From the observational perspective, this feature leads to the prediction that a black hole reaching a mass range between $10^9 - 10^{13}$g should create powerful short-lived gamma-ray bursts with energy of a few hundred MeV \cite{haw}. So far, all attempts to detect such high energy bursts have failed and resulted only in upper bounds on black hole evaporation rate in the vicinity of Earth \cite{gamma,gamma1}.

Thirdly, a cold phase in a black hole's life gives an interesting perspective on dark matter \cite{mc,rov,dm}. If the evaporation process is halted as some mass (potentially set by the Plack-mass) one may end up with a stable configuration which, except for its gravitational interaction, has no or extremely small interaction with ordinary matter and hence fits perfectly into the definition of a Weakly Interaction Massive Particle  (WIMP). This has led many authors to claim that in fact such tiny black holes could explain the mystery of dark matter in our universe, see, e.g., \cite{mc,rov,dm} for selected references. In reference \cite{revi} it is shown that no major constrain can be cast upon the properties of Planck-size remnants if they play the role of dark matter at a cosmological scale; nonetheless, the way these remnants can be produced and their stability could be potential weak spots of such scenarios \cite{rov}. 

These puzzles, clearly ask for a better understanding of the black hole evaporation process  beyond the quantum field theory in curved spacetime analysis. It is conceivable, that the ultimate answer lies in the realm of a theory of quantum gravity. Since there are currently many different routes at various stages of development, we take a different angle on the problem: quite strikingly many quantum gravity programs, including string theory
\cite{Becker:2006dvp,Zwiebach:2004tj,Palti:2019pca},
 loop quantum gravity and spin foams \cite{Ashtekar:2021kfp,Perez:2012wv,Rovelli:2014ssa,Ashtekar:2021kfp},
 asymptotically safe gravity \cite{Niedermaier:2006wt,Reuter:2012id,Percacci:2017fkn,Eichhorn:2018yfc,Reuter:2019byg,Pereira:2019dbn,Reichert:2020mja,Pawlowski:2020qer},
Causal Dynamical Triangulations \cite{Ambjorn:2012jv,Loll:2019rdj},
and Ho\v{r}ava-Lifshitz gravity \cite{Hor,Wang:2017brl,Barvinsky:2021ubv} predict a dynamical dimensional reduction of the theories momentum space \cite{Carlip:2017eud,Carlip:2019onx} (also see \cite{tHooft:1993dmi} for an early account of this idea). Specifically, the spectral dimension $D_s$, probing the effective dimension experienced by a random walk, drops from $D_s = 4$ at macroscopic scales to $D_s \approx 2$ at short distances. For instance, the analysis  of geometries obtained from the Causal Dynamical Triangulations program reported a scale-dependent spectral dimension \cite{Ambjorn:2005db},
\be
D_s(T) = a - \frac{b}{c+T} \, , 
\ee
where $a=4.02$, $b=119$, and $c =54$. A similar analysis within Euclidean Dynamical Triangulations \cite{Laiho:2017htj} obtained $D_s(T) = 3.94 \pm 0.16$ and $D_s(T) = 1.44 \pm 0.19$ for the large and small distance values when extrapolating to the continuum and infinite volume limits. Generalizing \cite{Lauscher:2005qz}, the scale-dependent spectral dimension along a fixed asymptotically safe renormalization group trajectory has been computed in \cite{Reuter:2011ah}, leading to a three plateau structure with $D_s(T) =4$, $D_s(T) = 4/3$, and $D_s(T)=2$ at large, intermediate, and short distances, respectively (also see \cite{Reuter:2012xf} for a review).

In \cite{Barcaroli:2015xda} the scale-dependence of the spectral dimension has been linked to the dimensionality of the theories momentum space: a drop of the spectral dimension indicates that there are less degrees of freedom at high-energy as compared to the expectation based on the dimension of spacetime observed at macroscopic scales. Thus the dynamical dimensional reduction provides a powerful mechanism for eliminating divergences occurring at high energies. Within the realm of multi-scale models \cite{cal,Calcagni:2013vsa,Calcagni:2016azd,Calcagni:2015xcf}, the phenomenological consequences of this mechanism have been explored, e.g., in the context of quantum field theory \cite{cal-standard}, cosmology \cite{cal-cosmo,cal-cosmo2}, and also for the Unruh effect \cite{Alkofer:2016utc}, see \cite{Calcagni:2021ipd} for an up-to-date review.\footnote{Along different lines, fractal aspects of black holes have been considered within the ``un-gravity program'' \cite{grav1,spec1}.}

In \cite{carlip} the authors suggested an intricate connection between dynamical dimensional reduction and the formation of cold remnants formed at the end of the black hole evaporation process based on a two-dimensional dilaton-gravity model. The goal of our work is to complement this analysis by implementing the effect of a drop in the spectral dimension in the thermodynamic properties of a four-dimensional Schwarzschild black hole. As our main result, we demonstrate that the mechanism of dynamical dimensional reduction removes the thunderbolt singularity appearing in the last stages of the black hole evaporation process. It does not lead to the formation of long-lived black hole remnants though. The latter requires additional ingredients, with a change in the topology of the black hole solution being the most probable one.

The rest of our work is organized as follows. Sect.\ \ref{sect.2} and Sect.\ \ref{sect.3} provide a brief introduction to black hole thermodynamics and the concept of generalized dimensions, respectively. Our analysis is presented in Sect.\ \ref{sect.4} and we conclude with a brief discussion and outlook in Sect.\ \ref{sect.5}.   

\section{Black hole thermodynamics in a nutshell} 
\label{sect.2}
We start by reviewing the basics of black hole thermodynamics, referring to \cite{Davies,Wald:1995yp,Raine:2005bs} for more detailed, pedagogical accounts. For simplicity, we consider spherically symmetric black holes described by the Schwarzschild solution. In natural units where $G=c=\hbar=k_b=1$, the resulting line-element is
\be\label{eq:sssol}
ds^2 = (1 - \frac{2M}{r}) dt^2 - (1 - \frac{2M}{r})^{-1} dr^2 - r^2 d\Omega^2  \, . 
\ee 
Here $d\Omega^2 = d\theta^2 + \sin^2\theta d\phi^2$ is the line-element on the unit two-sphere and $M$ is the mass of the black hole. The geometry \eqref{eq:sssol} possesses an event horizon at
\be
r_h = 2M \, . 
\ee
Based on \eqref{eq:sssol} one readily deduces that the area of this horizon is
\be\label{eq:horizonarea}
A_h = 4 \pi r_h^2 = 16 \pi M^2 \, . 
\ee
Any object or photon crossing this horizon inevitably has to move inward, eventually ending at the curvature singularity at $r=0$. Classically, signals emitted at $r \le r_h$ can not reach an observer stationed at $r > r_h$. Hence the terminology ``black hole''.

\begin{figure}[h] 
	\includegraphics[scale=0.60]{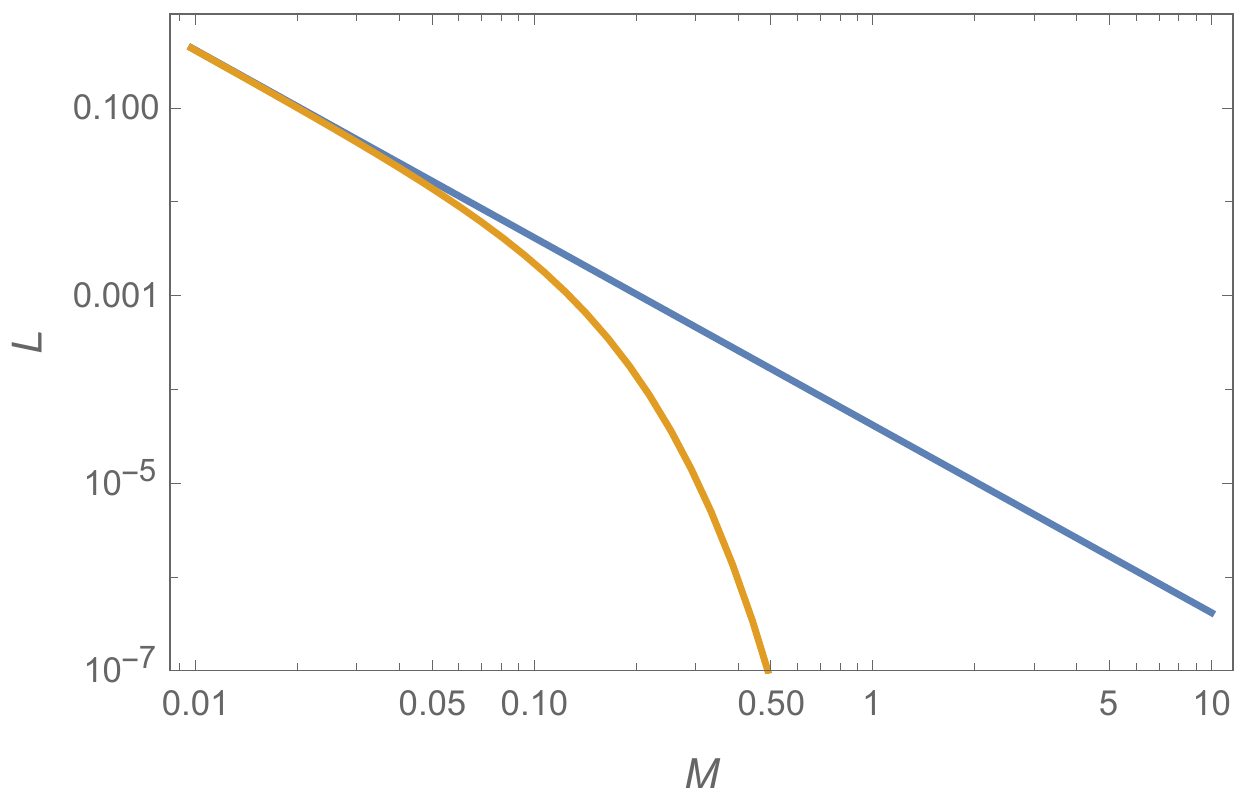} \\
	\caption{\label{Hlumin} Luminosity of a Schwarzschild black hole as a function of the mass $M$. The case of a massless and a massive scalar field with $m^2 = 1$ are illustrated by  the straight blue and orange lines, respectively.}
\end{figure}

The analysis within the framework of quantum field theory in curved spacetime \cite{Hawking:1975vcx} shows, however, that the event horizon emits black body radiation (Hawking radiation) with a temperature proportional to the surface gravity at the  horizon. For the Schwarzschild black hole \eqref{eq:sssol}, this results in
\be\label{eq:Hawkingtemp}
T_h = \frac{1}{8\pi M} \, . 
\ee
The resulting luminosity $L$ is then given by
\be\label{eq:Lgeneral}
L = A_h \, I \, , 
\ee
where $I$ is the integrated black body spectrum. For bosonic fields as the scalar field considered in this work
\be\label{eq:bbfactor}
I = \int \frac{d^3p}{(2\pi)^3} \frac{E}{e^{E/T_h} - 1} \, . 
\ee
For a massless photon $E = |\vec{p}| = \omega$ and one recovers the standard result \cite{Raine:2005bs}:\footnote{In general, the power contained in the Hawking radiation associated with a massless scalar field has the form $P = \sum_l \int_0^\infty d\omega P_l(\omega)$ with the $l$th multipole contributing with $P_l(\omega) = \frac{A_h}{8\pi^2} T_l(\omega) \omega^3 (e^{\omega/T_h}-1)^{-1}$. Our analysis focuses on the $l=0$ sector and neglects the gray-body corrections $T_l(\omega)$. Since the latter encode the transmission probability of Hawking radiation reaching future infinity without being backscattered by the gravitational barrier surrounding the black hole, one expects that these will lead to an additional suppression of the massive contributions as compared to the massless ones. Based on eq.\ \eqref{eq:Lspec} one then expects that the inclusion of the $T_l(\omega)$ will further inhibit the formation of remnants while leaving the leading order analysis unaffected.}
\begin{equation} \label{Lumin}
	\begin{split}
	L_{\rm massless} \,= & \, \frac{8 M^2}{\pi} \int_{0}^{\infty} d\omega \, \frac{\omega^3}{e^{8\, \pi\, M\, \omega}-1} \\
	= & \, \frac{1}{7680 \pi M^2} \, . 
	\end{split} 
\end{equation}
Here we have performed the integral over frequencies and expressed the result in terms of the black hole mass $M$ by substituting \eqref{eq:Hawkingtemp}. $L_{\rm massless}$ as a function of $M$ is then illustrated as the blue straight line in Fig.\ \ref{Hlumin}. Eq.\ \eqref{Lumin} exhibits the curious feature that black holes become more and more luminous the lighter they become. In particular $L$ diverges as $M \rightarrow 0$. The presence of this so-called thunderbolt singularity suggests that the semi-classical analysis breaks down when describing the final stage of black hole evaporation \cite{Hawking:1992ti,Piran:1993tq,Ashtekar:2010hx,Lowe:1993zw}.

The life-time of a black hole with initial mass $M_0$ can then be obtained by integrating the mass-loss formula
\be\label{eq:massloss}
\frac{d M}{dt} = - L \, . 
\ee
Subsituting \eqref{Lumin} gives the black hole evaporation time
\be
t_{\rm evap} = 2560 \, \pi \, M_0^3 \, . 
\ee
Thus the emission of Hawking radiation renders the life-time of the black hole finite.

The luminosity formula \eqref{eq:Lgeneral} is readily generalized to the case of a scalar field with mass $m$. In this case the energy appearing in \eqref{eq:bbfactor} is replaced by the relativistic dispersion relation $E^2 = \vec{p}^2 + m^2$. In general, $I$ does not admit an simple analytic expression. Nevertheless, it is instructive to study the following limits: if $m/T_h \ll 1$ the particle is relativistic and one essentially recovers the massless case. For temperatures $m/T_h \gg 1$ the particle is non-relativistic and the Bose-Einstein distribution in \eqref{eq:bbfactor} may be approximated by the Boltzmann distribution. This shows that the contribution of a massive mode is actually exponentially suppressed at temperatures $T_h \lesssim m$. The full expression for $L_{\rm massive}$ as a function of $M$ is obtained by numerical integration. The result is the orange line shown in Fig.\ \ref{Hlumin}.
  
\section{The Spectral Dimension in a nutshell} 
\label{sect.3}
An intuitive picture about quantum gravity is that spacetime at short distances will develop non-manifold like features. A first step towards characterizing the resulting structures is to generalize the notion of ``dimension'' borrowing concepts from fractal geometry \cite{book-diffusion}. In this way one naturally distinguishes between the Hausdorff dimension (based on covering a set of points with balls of decreasing radius), the spectral dimension (measuring the dimension ``felt'' by a diffusing particle), or the walk dimension (related to the expectation value of the distance traveled by a random walk as function of the diffusion time) of an Euclidean space. While all of these dimensions agree when working on manifolds, they characterize distinct properties of fractal spaces. In the present work, the key role is played by the spectral dimension $D_s$ which may be interpreted as the dimension of the theories momentum space \cite{Barcaroli:2015xda}. A decrease of the spectral dimension at high energy may then be interpreted as ``the theory possessing less degrees of freedom than its analogue defined on a background manifold''. 

Formally, the spectral dimension $d_s$ and its scale-dependent generalization $D_s(T)$ is introduced by studying the diffusion of a test particle on a $d$-dimensional Euclidean spacetime with metric $g_{\mu\nu}$ with respect to the fiducial diffusion time $T$. Denoting the Laplacian constructed from $g_{\mu\nu}$ by $\Delta \equiv - g^{\mu\nu}D_\mu D_\nu$, and introducing $F(\Delta) \equiv G(\Delta)^{-1}$ with $G(\Delta)$, being the position-space representation of the particles propagator, the motion of the test particle is captured by the generalized heat equation
\begin{equation}\label{diff}
	\frac{\partial}{\partial T} K_{g}(\xi,\xi_0;T)= -F(\Delta) K_{g}(\xi,\xi_0;T)
\end{equation} 
subject to the boundary condition
\be
K_{g}(\xi,\xi_0;T)|_{T = 0} = \delta^d(\xi - \xi_0) \, . 
\ee
Here $K_{g}(\xi,\xi_0;\sigma)$ is the heat-kernel associated with $F(\Delta)$. It describes the probability of the particle defusing from the initial point $\xi_0$ to $\xi$ during the time-interval $T$. In particular, one recovers the standard heat-equation for $F(\Delta) = \Delta$. The return probability $P_g(T)$ is then defined by the particle returning to its initial point after time $T$
\be\label{eq:return}
 P_g(T) \equiv V^{-1} \int d^d\xi \sqrt{g} \, K_{g}(\xi,\xi;T) \, . 
\ee
Here $V \equiv \int d^d\xi \sqrt{g}$ is the volume of the space. Based on \eqref{eq:return}
the spectral dimension $d_s$ is then defined as
\be\label{def:specdim}
d_s \equiv -2 \lim_{T \rightarrow 0} \, \frac{d \ln P_g(T)}{d \ln T} \, . 
\ee
For the standard heat-equation on a smooth manifold $d_s = d$ agrees with the topological dimension of the manifold. At this stage, it is convenient to generalize \eqref{def:specdim}, allowing for a scale-dependent spectral dimension
\begin{equation}\label{def:specdimT}
	D_s(T) \equiv -2\frac{d \ln P_g(T)}{d \ln T} \, . 
\end{equation}
$D_s(T)$ takes into account the possibility that long random walks my experience a different spectral dimension than the infinitesimal ones entering in the definition \eqref{def:specdim}.

On a flat Euclidean space $\mathbb{R}^d$ with metric $\delta_{\mu\nu}$ the generalized heat-equation \eqref{diff} can be solved using Fourier-techniques
\begin{equation}\label{eq:diffkernel}
	K_{\delta}(\xi,\xi_0;T)=\int \frac{d^dp}{(2\pi)^d}e^{ip(\xi-\xi_0)} e^{-T F(p^2)} \, . 
\end{equation}
The resulting return probability is
\begin{equation}\label{eq:retprob}
	P_\delta(T)=\int \frac{d^dp}{(2\pi)^d} e^{-T F(p^2)} \, . 
\end{equation}
For $F(p^2) = p^2$, the return probability evaluates to
\be
P_\delta(T)= (4 \pi T)^{-d/2} \, . 
\ee
Substituting this result into \eqref{def:specdimT} shows that $D_s(T) = d$ is independent of $T$ and agrees with the topological dimension $d$. 

The computation is readily generalized to the case where the function $F(p^2)$ has a fixed scaling behavior $F(p^2) = p^{2+\delta}$.\footnote{Generically, any function $F(p^2)$ for which the integral \eqref{eq:diffkernel} is not the Fourier-transform of a Gaussian will result in diffusion kernels $K_g(\xi,\xi_0;T)$ which are not positive semi-definite. The occurrence of negative probabilities can be cured by going to fractional calculus \cite{cal}. Since this is not relevant in the present analysis, we do not dwell on this technical feature at this point.}
 Rewriting the integral in \eqref{eq:retprob} in terms of the dimensionless variable $x = p^{2+\delta} T$, one readily finds that \cite{Reuter:2011ah}
\begin{equation}\label{eq:dsT}
	D_s(T)=\frac{2d}{2+\delta} \, , 
\end{equation}
which is again independent of the diffusion time $T$.

\begin{figure}[h] 
	\includegraphics[scale=0.60]{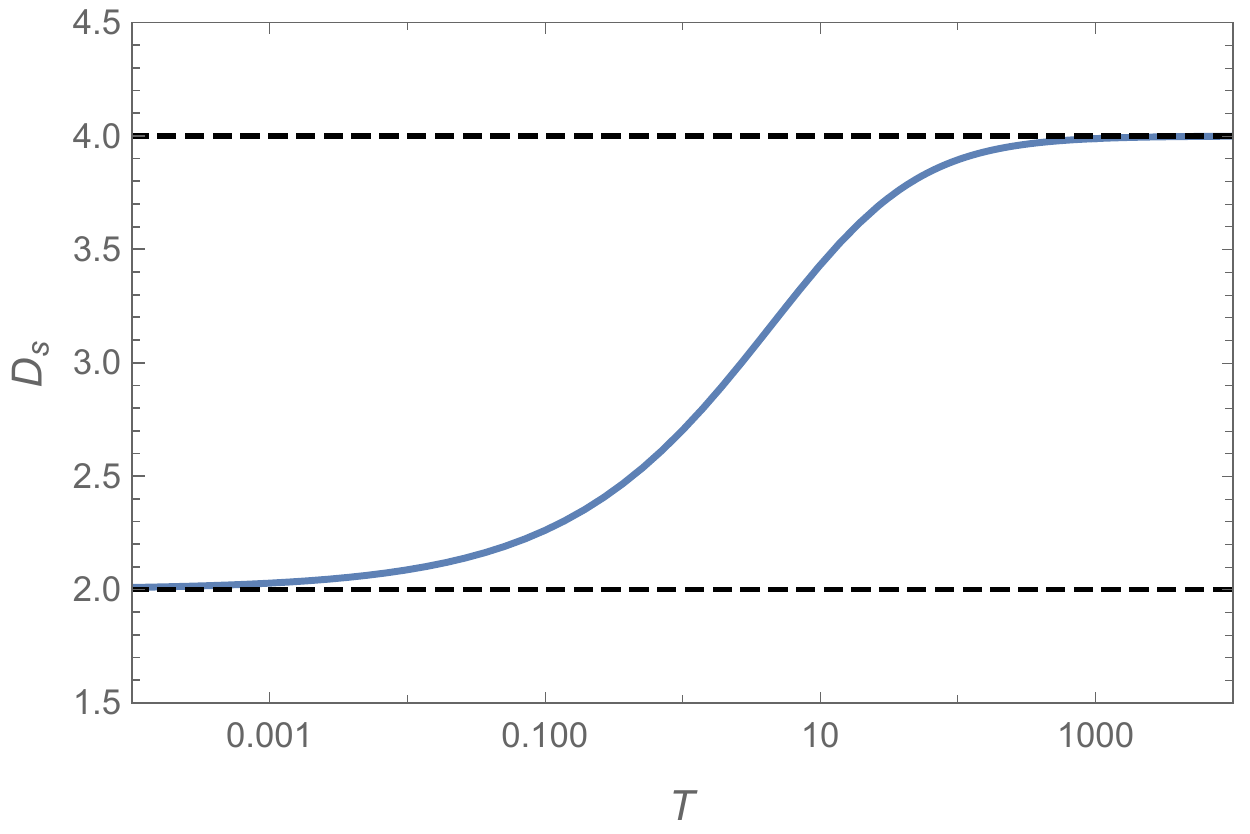}
	\caption{\label{crossover} Illustration of the scale-dependent spectral dimension $D_s(T)$ obtained from the two-scale model \eqref{mod} with $m^2 =1$. $D_s(T)$ interpolates smoothly between $D_s = 4$ for $T/m \gg 1$ and $D_s = 2$ for $T/m \ll 1$.}
\end{figure}

Based on \eqref{eq:dsT} it is then straightforward to construct a simple multi-scale model which interpolates between $D_s(T) = 2$ at microscopic and $D_s(T) = 4$ at macroscopic scales \cite{Alkofer:2016utc}. Starting from the momentum-space propagator
\begin{equation}\label{mod}
	G(p^2)=\frac{1}{p^2}-\frac{1}{p^2+m^2} \, , 
\end{equation}
one obtains $F(p^2) = \frac{1}{m^2} \, p^2 (p^2 + m^2)$. Thus $F(p^2)$ interpolates between $F(p^2) \propto p^2$ for $p^2 \ll m^2$ and $F(p^2) \propto p^4$ for $p^2 \gg m^2$. Evaluating \eqref{eq:dsT} in these scaling regimes suggests that one recovers the desired behavior of the spectral dimension at microscopic and macroscopic scales. The integrals determining $P_\delta(T)$ can be performed analytically and can be expressed in terms of error-functions. The resulting spectral dimension is shown in Fig.\ \ref{crossover}. This confirms that the model indeed interpolates between $D_s = 4$ for $T/m \gg 1$ and $D_s = 2$ for $T/m \ll 1$ 
Eq.\ \eqref{mod} then defines a generic toy model realizing the dynamical dimensional reduction encountered in the full-fledged quantum gravity analysis. The latter may also fix the cross-over scale $m$ based on microscopic considerations.

\section{Black hole thermodynamics including a non-trivial spectral dimension} 
\label{sect.4}
At this stage we are in a position to combine our discussions on the thermodynamical properties of black holes and dynamical dimensional reduction. Throughout this section we will assume that the radiation emitted by the event horizon remains thermal also for very light black holes, see \cite{thermality,thermality1,thermality2} for a detailed analysis supporting this assumption. The goal of this section is then to go beyond the semi-classical analysis utilizing the concepts of generalized dimensions and dynamical dimensional reduction.
\subsection{Single-scale analysis}
\label{sect.4.1}
From the perspective of generalized dimensions the luminosity formula \eqref{eq:Lgeneral} contains two distinguished elements. Firstly, the black body factor $I$ contains an integral over the theories momentum space. This suggests that the dimension appearing in this term is the spectral dimension
\be\label{eq:Ispec}
I_{d_s} = \int \frac{d^{d_s-1}p}{(2\pi)^{d_s-1}} \frac{E}{e^{E/T_h} - 1} \, . 
\ee
Secondly, the horizon area is related to position space properties. This suggests that this term is sensitive to the Hausdorff dimension of the (quantum) spacetime. Owed to the lack of a concrete model which would allow to describe such an effect for the event horizon, we refrain from including such an effect. If a concrete model is available, this feature may be included rather straightforwardly in the present setting by considering an effective dimension build from a linear combination of the spectral and Hausdorff dimension.

The generalization \eqref{eq:Ispec} then allows to determine the threshold on the spectral dimension $d_s$ required for creating a long-lived black hole remnant from dynamical dimensional reduction in the momentum space. For this purpose we consider \eqref{eq:Ispec} for a massless scalar field in a scaling regime where $d_s$ is constant but not necessary identical to the topological dimension $d$ of the spacetime. Convergence of the integral requires $d_s > 1$. Assuming convergence, $I_{d_s} = \tilde{c} M^{-d_s}$ where $\tilde{c}$ is a numerical constant. In combination with the classical horizon area \eqref{eq:horizonarea} the luminosity obtained from the spectral dimension is
\be
L_{d_s} = 16 \pi \tilde{c} \, M^{2-d_s} \, . 
\ee
The mass-loss formula \eqref{eq:massloss} then shows that the generation of a remnant for which $t_{\rm evap}$ is infinite requires $d_s-2 \le -1$ or, equivalently, $d_s \le 1$. This is, however, in conflict with requiring convergence of $I_{d_s}$. Thus just modifying the spectral dimension for the fields constituting the Hawking radiation is not sufficient to create a long-lived remnant. In particular, the black hole evaporation does not stop if $d_s$ drops below three. 
\subsection{Dynamical dimensional reduction}
\label{sect.4.2}
Notably, the luminosity \eqref{eq:Lgeneral} \emph{is linear} in the two-point correlation function of the corresponding fields. The Euclidean multi-scale model following from \eqref{mod} then suggests the following generalization to the black hole context. First, the Euclidean flat-space propagators are analytically continued to Lorentzian signature using a standard Wick rotation. Subsequently, the principle of covariance is used to promote the derivatives appearing in the position space representation to covariant derivatives. In this way one naturally arrives at the conclusion that the luminosity $L_{D_s}$ of a black hole, in a situation where the scalar field modeling the Hawking radiation exhibits dynamical dimensional reduction, is given by the sum of a massless and massive contribution weighted by a relative minus sign:
\be\label{eq:Lspec}
L_{D_s}(M;m) = L_{\rm massless}(M) - L_{\rm massive}(M;m) \, . 
\ee
\begin{figure}[h] 
	\includegraphics[scale=0.60]{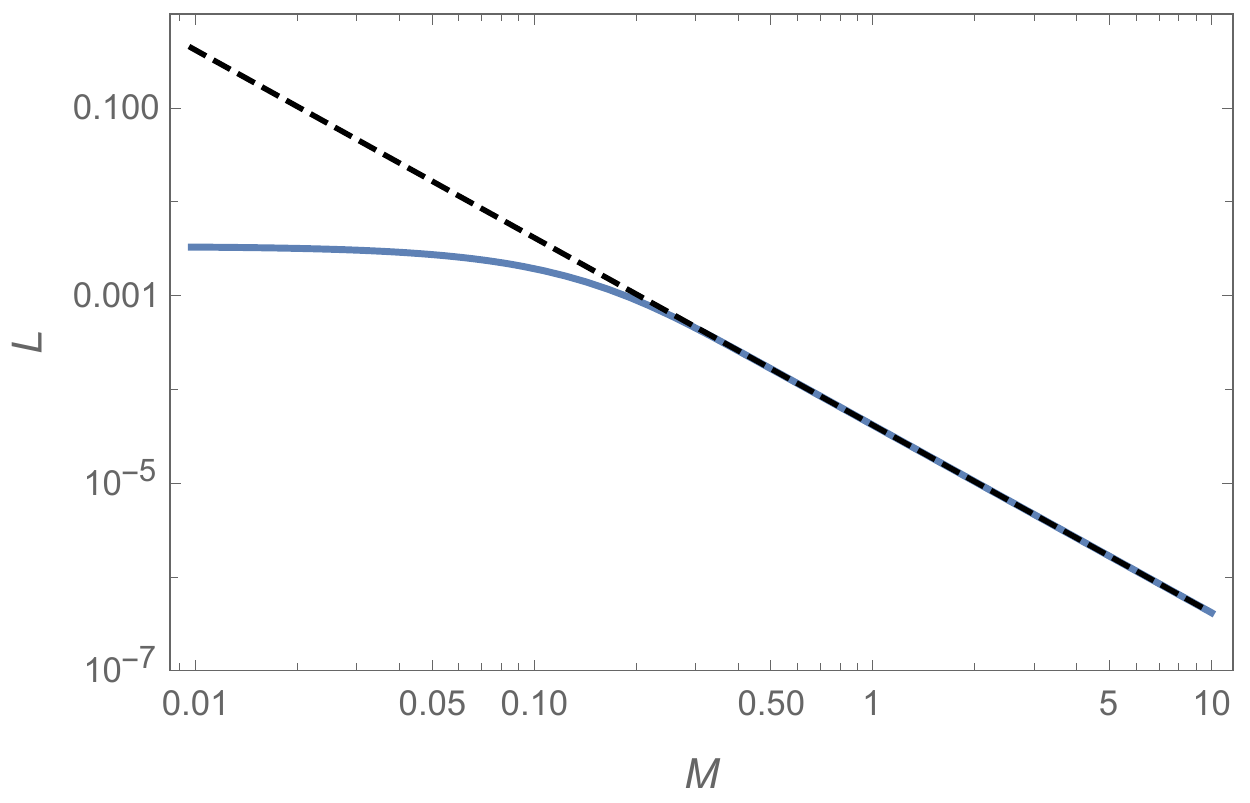}
	\caption{\label{luminds} Luminosity $L_{D_s}$ of a Schwarzschild black hole with mass $M$ arising from the two-scale model \eqref{mod} with $m^2 = 1$ (blue line). The inclusion of the massive mode triggering the dynamical dimensional reduction renders the luminosity finite as $M \rightarrow 0$. The massless case is added as the dashed line for reference.}
\end{figure} 
From the general analysis in Sect.\ \ref{sect.2} we then conclude that the contribution of the second term is exponentially suppresses for $M \gg m$. As a result, $L_{D_s}(M;m)$ agrees with the semi-classical analysis in this regime. Conversely, for $M \lesssim m$ both terms contribute with equal magnitude. As a result $L_{D_s}$ remains finite as $M \rightarrow 0$. The crossover between these two regimes together with the removal of the thunderbolt singularity is illustrated in Fig.\ \ref{luminds}, where $L_{D_s}$ has been evaluated numerically for $m^2=1$.

The taming of the luminosity for light black holes arises from the interplay of the massless and massive degree of freedom which provides a Pauli-Villars-type regularization for $L_{\rm massless}$. We stress that, from a quantum gravity perspective, this feature does not result from introducing an additional ghost field, but is a direct result of the reduced number of degrees of freedom exhibited by the theory at scales $|\vec{p}| \gtrsim m$.  

The result shown in Fig.\ \ref{luminds} together with a detailed numerical analysis reveals that $L_{D_s}(M;m)$ can be very well approximated by a simple interpolating function
\be\label{eq:Lspecana}
L_{D_s}(M;m) = \frac{c}{b \, m^{-2}  + M^2}
\ee
where
\be
c = \frac{1}{7680 \pi} \, , \qquad b \approx 0.0125 \, . 
\ee
The value for $b$ has been obtained from fitting the ansatz \eqref{eq:Lspecana} to $L_{D_s}(M,m)$, obtained via numerical integration, at values $M \ll 1$.  

The analytic formula \eqref{eq:Lspecana} again allows to compute the Hawking evaporation time of the black hole analytically. Integrating eq.\ \eqref{eq:massloss} yields
\be\label{eq:tevapspec}
t_{\rm evap} = 2560 \pi M_0^3 + \frac{b}{c} \frac{M_0}{m^2} \, . 
\ee
Thus the dynamical dimensional reduction leads to an increase of the black hole lifetime. Since the scale $m$ where the dynamical dimensional reduction sets in is expected to be the Planck scale, this is a rather tiny effect though. Evaluating the term linear in $M_0$ for $m = m_{\rm Planck} = 2.18 \times 10^{-5}$ g and $M_0 = 10^9$ g yields that the change in the lifetime of the black hole is given by $\Delta t_{\rm evap} = 7.46 \times 10^{-28}$ s. Thus the structure of \eqref{eq:tevapspec} indicates that a luminosity which is constant as $M \rightarrow 0$ does not lead to a long-lived remnant with a life-time comparable to cosmic time-scales.

\section{Conclusions and Discussion} 
\label{sect.5}
Motivated by the observation that light black holes with a mass given by the Planck mass $M_{\rm Pl} \approx 10^{-5}$g may constitute valid dark matter candidates \cite{mc,rov,dm,mur}, we used black hole thermodynamics to investigate the luminosity and lifetime of spherically symmetric black hole solutions. Our work stepped out of the perturbative framework of quantum field theory in curved spacetime by including the effect of a dynamical dimensional reduction of the theories momentum space. As this is a feature shared by many approaches to quantum gravity \cite{Carlip:2017eud,Carlip:2019onx}, it is intriguing to investigate whether this mechanism leads to the formation of long-lived black hole remnants. While we showed that the dynamical dimensional reduction mechanism generically removes the divergences in the black hole luminosity encountered in the last stage of the evaporation process, the results do not provide any evidence supporting the formation of long-lived remnants. 

At this point the following remarks on the scope and limitations of our analysis are in order. Our work implemented the mechanism of dynamical dimensional reduction at the level of the degrees of freedom constituting the Hawking radiation. In this course \emph{we did not modify the topology of the background black hole solution} which is given by the Schwarzschild solution. This entails that our spacetimes exhibit just one horizon, the event horizon and there is no inner horizon. This feature is at variance with many proposals for quantum gravity inspired black hole metrics as, e.g., the Hayward metric \cite{Hayward}, the renormalization group improved black hole solutions constructed by Bonanno and Reuter \cite{martin}, or the Planck stars inspired by Loop Quantum Gravity \cite{loop-rem}. A direct consequence of our topology is that the black holes in our work do not have a critical mass where the two horizons coincide and the surface gravity (and hence the Hawking temperature) is zero. By disentangling the effects of dynamical dimensional reduction and the topology of spacetime, our analysis clearly reveals that it is the latter ingredient which is decisive for forming a light black hole remnant  during the final stages of black hole evaporation. 

This observation is also the key for reconciling our results with the ones reported by Carlip and Grummiller \cite{carlip}. In this case the effect of dynamical dimensional reduction was essentially incorporated through generalizing the scaling law for the event horizon,
\be\label{eq:eh}
A_h = 4 \pi r_h^{d_h - 2} \, , 
\ee
and subsequently identifying $d_h = d_s$. Within the single-scale analysis of Sect.\ \ref{sect.4.1} this modification with $d_h = 2$ or $d_h = 3$ would not lead to a stop of the Hawking evaporation process. A careful analysis of the dilaton model used in \cite{carlip} shows, however, that the underlying black hole solutions must come with a second horizon: in this way one can approach a critical configuration if the generalized dimension of the dilaton model $D(X) = 3$. This picture then corroborates our conclusion that it is actually the topology of the (quantum) black hole and not the effect of a dynamical dimensional reduction that is crucial for forming light long-lived remnants.

Naturally, it would be interesting to base our conclusion on a first-principle derivation from a full-fledged theory of quantum gravity. Such a derivation will require detailed knowledge about the momentum-dependence of the theories two-point functions. Notably, the form factor program for Asymptotic Safety \cite{Knorr:2019atm} has recently made substantial progress along these lines \cite{Knorr:2018kog,Bosma:2019aiu,Draper:2020bop,Platania:2020knd,Bonanno:2021squ,Knorr:2021niv}. Clearly, it would then be interesting to evaluate the black hole luminosity as an observable sensitive to a non-trivial momentum dependence in the propagators of the fields. We hope to come back to this point in the future.

\begin{acknowledgments}
	We thank C.\ Laporte for interesting discussions. The work by F.S.\ is supported by the NWA-grant ``The Dutch Black Hole Consortium''.
\end{acknowledgments}

\end{document}